\documentclass[%
reprint,
superscriptaddress,
amsmath,amssymb,
aps,
pra,
]{revtex4-2}
\usepackage{ragged2e}
\renewcommand{\justify}{\leftskip=0pt \rightskip=0pt plus 0cm}
\usepackage{floatrow}
\usepackage{graphicx}
\usepackage{subfigure}
\usepackage{filecontents}
\usepackage{pgfplots}
\usepackage{amsmath}
\usepackage{amssymb}
\usepackage{tikz}
\usepackage{dcolumn}
\usepackage{bm}
\usepackage{blindtext}
\usepackage{listings}
\usepackage{physics}
\usepackage{stmaryrd}
\usetikzlibrary{quantikz}
\usepackage[colorlinks,linkcolor=blue,anchorcolor=blue,citecolor=blue]{hyperref}
\makeatletter
\newif\if@restonecol
\makeatother

\usepackage[ruled,linesnumbered,vlined]{algorithm2e}
\usepackage{algpseudocode}

\begin{document}
\title{Quantum error correction with the color-Gottesman-Kitaev-Preskill code}

\author{Jiaxuan Zhang}
\affiliation{Key Laboratory of Quantum Information, Chinese Academy of Sciences, School of Physics, University of Science and Technology of China, Hefei, Anhui, 230026, P. R. China}
\affiliation{CAS Center For Excellence in Quantum Information and Quantum Physics, University of Science and Technology of China, Hefei, Anhui, 230026, P. R. China}

\author{Jian Zhao}
\affiliation{Key Laboratory of Quantum Information, Chinese Academy of Sciences, School of Physics, University of Science and Technology of China, Hefei, Anhui, 230026, P. R. China}
\affiliation{CAS Center For Excellence in Quantum Information and Quantum Physics, University of Science and Technology of China, Hefei, Anhui, 230026, P. R. China}

\author{Yu-Chun Wu}
\email{wuyuchun@ustc.edu.cn}
\affiliation{Key Laboratory of Quantum Information, Chinese Academy of Sciences, School of Physics, University of Science and Technology of China, Hefei, Anhui, 230026, P. R. China}
\affiliation{CAS Center For Excellence in Quantum Information and Quantum Physics, University of Science and Technology of China, Hefei, Anhui, 230026, P. R. China}
\affiliation{Institute of Artificial Intelligence, Hefei Comprehensive National Science Center, Hefei, Anhui, 230088, P. R. China}

\author{Guo-Ping Guo}
\affiliation{Key Laboratory of Quantum Information, Chinese Academy of Sciences, School of Physics, University of Science and Technology of China, Hefei, Anhui, 230026, P. R. China}
\affiliation{CAS Center For Excellence in Quantum Information and Quantum Physics, University of Science and Technology of China, Hefei, Anhui, 230026, P. R. China}
\affiliation{Institute of Artificial Intelligence, Hefei Comprehensive National Science Center, Hefei, Anhui, 230088, P. R. China}
\affiliation{Origin Quantum Computing Hefei, Anhui 230026, P. R. China}
\date{\today}
\begin{abstract}
The Gottesman-Kitaev-Preskill (GKP) code is an important type of bosonic quantum error-correcting code. Since the GKP code only protects against small shift errors in $\hat{p}$ and $\hat{q}$ quadratures, it is necessary to concatenate the GKP code with a stabilizer code for the larger error correction. In this paper, we consider the concatenation of the single-mode GKP code with the two-dimension (2D) color code (color-GKP code)  on the square-octagon lattice. We use the Steane type scheme with a maximum-likelihood estimation (ME-Steane scheme) for GKP error correction and show its advantage for the concatenation. In our main work, the minimum-weight perfect matching (MWPM) algorithm is applied  to decode the color-GKP code. Complemented with the continuous-variable information from the GKP code, the threshold of  2D color code is improved. If only data GKP qubits are noisy, the threshold reaches $\sigma\approx 0.59$ $(\bar{p}\approx13.3\%)$ compared with $\bar{p}=10.2\%$ of the normal 2D color code. If measurements are also noisy, we introduce the generalized Restriction Decoder on the three-dimension space-time graph for decoding.  The threshold reaches $\sigma\approx 0.46$ when measurements in the GKP error correction are noiseless, and $\sigma\approx 0.24$ when all measurements are noisy. Lastly, the good performance of the generalized Restriction Decoder is also shown  on the normal 2D color code giving the threshold at $3.1\%$ under the phenomenological error model. 
\end{abstract}

\maketitle
\section{Introduction}

Quantum error correction is one of the fundamental requirements for  practical large-scale quantum computation \cite{terhal2015quantum,gottesman1997stabilizer,preskill1998reliable}. The main idea of quantum error correction is encoding quantum information in a large quantum system. One effective way is storing redundant information in qubit-based quantum systems where logical qubits are encoded by a large amount of physical qubits. Many experiments have demonstrated quantum error correction with multiple qubits in various platforms such as superconducting \cite{reed2012realization,
kelly2015state,corcoles2015demonstration,takita2017experimental,
andersen2020repeated,gong2019experimental,
ai2021exponential}, ion traps \cite{nigg2014quantum,
egan2020fault,ryan2021realization} and nuclear magnetic resonance (NMR)  \cite{cory1998experimental,
moussa2011demonstration}  systems.  However, scaling up the number of qubits in these systems is extremely challenging \cite{corcoles2019challenges}. 

A popular alternative way for quantum error correction is the bosonic error correction code \cite{albert2018performance,cai2021bosonic}. In the bosonic architecture, the logical qubits are encoded in a continuous-variable system, i.e., bosonic modes. Bosonic modes  provide infinite-dimensional Hilbert space for quantum information encoding. The representative bosonic codes based on a single bosonic mode include the cat code \cite{ofek2016extending}, binomial code \cite{hu2019quantum} and Gottesman-Kitaev-Preskill (GKP) code \cite{gottesman2001encoding,grimsmo2021quantum}.

Recently, with the development of quantum hardware technology, the GKP code  has aroused extensive attention to realize practical bosonic error correction \cite{campagne2020quantum}. There are three main options for the GKP code error correction, the Steane type scheme \cite{steane1996error,gottesman2001encoding,terhal2016encoding}, the Knill-Glancy type scheme \cite{glancy2006error,wan2020memory} and the teleportation-based scheme \cite{walshe2020continuous,noh2021low}, in which the Steane type error correction scheme is most widely mentioned. However, when the ancilla qubits are noisy, the conventional Steane scheme does not provide the optimal solution for the error correction. Consider an example where the data qubit and the ancilla qubit have the error shift $u_1$ and $u_2$ respectively. Suppose the variances of  $u_1$ and $u_2$ are equal, the conventional Steane scheme applies the correction $q_{cor}=(u_1+u_2)\,{\rm mod} \sqrt{\pi}$ in data qubits. A simple analysis shows that replacing $u_1$ with $-u_2$ contributes nothing to the error correction, as $u_1$ and $u_2$ have equal variances. Actually, as noted in Section \ref{pre}, the Steane type error correction scheme with a maximum-likelihood estimation (ME-Steane scheme) \cite{fukui2019high} gives a better shift correction $q^{\scriptscriptstyle (ME)}_{cor}=\frac{1}{2}[(u_1+u_2)\,{\rm mod} \sqrt{\pi}]$ in this special case.

Since the GKP code error correction schemes are only used to correct small shift errors in the position and momentum quadratures of an oscillator, it is required to concatenate the GKP code with a stabilizer code to protect against larger errors \cite{menicucci2014fault}. For instance, the concatenation of GKP code with the toric code \cite{vuillot2019quantum} or surface code \cite{yamasaki2020polylog,noh2020fault,noh2021low} has been proposed, some of which is analyzed thoroughly under the circuit-level error model \cite{
noh2020fault,noh2021low}.

This paper is intended to study the GKP code concatenated with another topological stabilizer code -- color code \cite{bombin2006topological,bombin2007exact}. It is well-known that the logical Clifford group \cite{gottesman1998heisenberg} can be implemented transversally in color code \cite{fowler2011two} which is deemed to be a merit in the competition with the surface code. However, decoding the color code is generally considered much harder than the toric code or surface code since it
requires hypergraph matching, for which efficient algorithms are
 unknown \cite{wang2009graphical}. Several efficient decoders reveal the threshold of 2D color code around $8\%$ \cite{sarvepalli2012efficient,delfosse2014decoding}, that is obviously below the threshold of 2D toric code over $10\%$ \cite{dennis2002topological}. Fortunately, the Restriction Decoder shows good performance in 8,8,4 color code (i.e., 2D color code on the square-octagon lattice) with the threshold at $10.2\%$ \cite
{kubica2019efficient}.
This gives us a brilliant prospect to decode the color-GKP code by an efficient algorithm.

In this paper, we first decode the color-GKP code under perfect measurements by using the Restriction Decoder \cite{kubica2019efficient,chamberland2020triangular} with the continuous-variable GKP information, and find the threshold improved from $\bar{p}\approx 10.2\%$ to $13.3\%$ $(\sigma \approx 0.59)$. Secondly, to decode color code with noisy measurements, the generalized Restriction Decoder is introduced. More specifically, the minimum-weight perfect matching (MWPM) algorithm is applied in the 3D space-time graph. When data qubits and stabilizer measurements are noisy but GKP error correction is perfect, the threshold reaches $\sigma\approx 0.46$, and when GKP error correction is also noisy, the threshold reaches $\sigma\approx 0.24$. These results show the performance of the color-GKP code is on par with the toric-GKP code under the same error models. We accredit this to the good performance of the generalized Restriction Decoder. To support this consequence, the generalized Restriction Decoder is used to decode the normal 2D color code, and shows the threshold at $3.1\%$ under the phenomenological error model. The threshold compares quite well to the result given by some low efficient decoders \cite{landahl2011fault}.

The rest of the paper is organized as follows. Section \ref{s2} starts with  reviewing some basic aspects of GKP error correction code and introduces color-GKP  code on the square-octagon lattice. Section \ref{pre} discusses  ME-Steane type GKP error correction scheme and compares it with the conventional scheme. Section \ref{s3} states two error models and  the decoding strategies in  detail. The numerical simulation results of decoding color-GKP  code are presented in Section \ref{s4} where we  compare our results with previous work. Lastly,  the conclusion of this paper and the outlook for future work are described in Section \ref{s5}.

\section{The color-GKP code}\label{s2}

In this section, we introduce the color-GKP code, i.e., the single-mode GKP code concatenated with the color code. In the first layer of concatenation, the GKP code encodes  qubits into the Hilbert space of harmonic oscillators. In the second layer, the standard square GKP qubits are used as the data qubits of the 2d color code. Our work only considers the color-GKP code on the square-octagon lattice, since the 8,8,4 color code has a high threshold using a efficient decoding algorithm \cite{kubica2019efficient}.

\subsection{The GKP error correction code}
The GKP error correction code was firstly proposed by Gottesman, Kitaev, and Preskill in 2001 \cite{gottesman2001encoding}, which encodes a qubit into a harmonic oscillator. For a harmonic oscillator,  the position and momentum operators are defined as:
\begin{equation}
\hat{q}=\frac{1}{\sqrt{2}}(\hat{a}+\hat{a}^\dag),\quad
\hat{p}=\frac{i}{\sqrt{2}}(\hat{a}-\hat{a}^\dag),
\end{equation}
where $\hat{a}$ and $\hat{a}^\dag$ are annihilation and creation operators satisfying $[\hat{a},\hat{a}^\dag]=1$. The code space is stabilized by two commuting stabilizer operators 
\begin{equation}
\hat{S_p}=e^{-i2\sqrt{\pi}\hat{p}},\quad
\hat{S_q}= e^{i2\sqrt{\pi}\hat{q}},
\end{equation}
which can be regarded as $2\sqrt{\pi}$ displacement in the $\hat{q}$ and $\hat{p}$ quadratures respectively. Then  the logical Pauli operators are:
\begin{equation}
\bar{X}=e^{-i\sqrt{\pi}\hat{p}},\quad
\bar{Z}= e^{i\sqrt{\pi}\hat{q}}.
\end{equation}
One can easily check that $\bar{X}$ (or $\bar{Z}$) commutes with $\hat{S_p}$ and $\hat{S_q}$, and satisfies $\bar{X}^2=\hat{S_q}$ and $\bar{Z}^2=\hat{S_p}$. Ideally, the logical states can be written as 
\begin{equation}\label{eq4}
\begin{aligned}
&|\bar{0}\rangle \propto 
\sum_{n\in \mathbb{Z}}\delta(q-2n\sqrt{\pi}) |{q}\rangle
=\sum_{n\in \mathbb{Z}} |{q}=2n\sqrt{\pi} \rangle,\\
&|\bar{1}\rangle \propto 
\sum_{n\in \mathbb{Z}}\delta(q-(2n+1)\sqrt{\pi}) |{q}\rangle
=\sum_{n\in \mathbb{Z}} |{q}=(2n+1)\sqrt{\pi} \rangle .
\end{aligned}
\end{equation}

The most relevant quantum gates to the quantum error correction are Clifford gates, the generators of which are Hadamard gate $H$, phase gate $S$, and control-not gate ${\rm{CNOT}}_{ij}$ in control qubit $i$ and target qubit $j$.  The Clifford gates of the GKP code can be performed by interactions that are at most quadratic in the creation and annihilation operators \cite{grimsmo2021quantum}, therefore $H$, $S$ and $\mathrm{CNOT}_{ij}$  have the following forms:
\begin{equation}
\begin{aligned}
&H=e^{i\pi\hat{a}^\dag\hat{a}/2},\quad S=e^{i\hat{q}^2/2},\\
&\mathrm{CNOT}_{ij} =e^{-i\hat{q_i}\hat{p}_j}.
\end{aligned}
\end{equation}

However, the logical states defined in Eq.$\,$(\ref{eq4}) are not physical, since these ideal states are not normalizable and require infinite energy to squeeze. In the realistic situation, $\delta$ functions in Eq.$\,$(\ref{eq4}) will be replaced by finitely squeezed Gaussian states weighted by a Gaussian envelope.  Therefore the noisy GKP states become
\cite{wang2019quantum}:
\begin{equation}
\begin{aligned}
&|\tilde{0}\rangle \propto \sum_{n\in \mathbb{Z}}   \int_{-\infty}^\infty
e^{-\frac{\Delta^2}{2}(2n)^2\pi}e^{{-\frac{1}{2\Delta^2}({q}-2n\sqrt{\pi})^2}}
|{q}\rangle dq,\\
&|\tilde{1}\rangle \propto \sum_{n\in \mathbb{Z}}   \int_{-\infty}^\infty
e^{-\frac{\Delta^2}{2}(2n+1)^2\pi}e^{{-\frac{1}{2\Delta^2}({q}-(2n+1)\sqrt{\pi})^2}}
|{q}\rangle dq.
\end{aligned}
\end{equation}

An arbitrary noisy GKP state $|\tilde{\psi}\rangle=\alpha|\tilde{0}\rangle+\beta|\tilde{1}\rangle$ can be regarded as the ideal GKP state
$|\bar{\psi}\rangle$ suffering the position and momentum shift errors with the Gaussian distribution:
\begin{equation}\label{eq7}
\begin{aligned}
|\tilde{\psi}\rangle\propto\iint  e^{-\frac{u^2+v^2}{2\Delta^2}} e^{-iu\hat{p}}e^{iv\hat{q}} |\bar{\psi}\rangle du dv.
\end{aligned}
\end{equation}
The proof of Eq.$\,$(\ref{eq7}) is shown in Appendix \ref{aa}.

\begin{figure}[t]
\centering
\includegraphics[width=8cm]{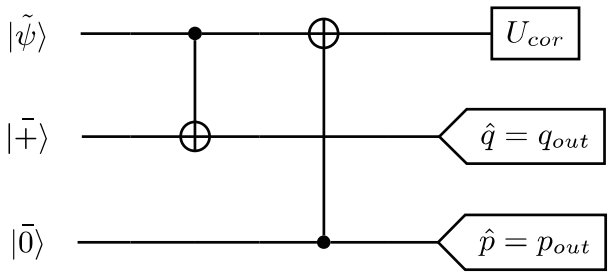} 
\caption{\justify Quantum circuit of the Steane type error correction. The Steane type error correction scheme can correct small shift errors in the $\hat{q}$ and $\hat{p}$ quadratures and requires two high-quality ancilla GKP qubits. Here we suppose $\bar{\ket{+}}$ and $\bar{\ket{0}}$ are the ideal GKP states and $\tilde{\ket{\psi}}$ is the noisy GKP state. The correction operator $U_{cor}=\exp(iq_{cor}\hat{p})\exp(ip_{cor}\hat{q})$, where $q_{cor}=$ $q_{out}\,{\rm mod}\sqrt{\pi}$ and $p_{cor}=p_{out}\,{\rm mod}\sqrt{\pi}$.}\label{fig1}
\end{figure}

By utilizing  Pauli twirling approximation \cite{katabarwa2017dynamical}, these errors  can be viewed as coming from a Gaussian shift error channel $ \mathcal{N}$ with variance $\sigma^2$ acting on the ideal GKP state:
\begin{equation}\label{ee8}
 \mathcal{N}(\rho)\equiv \iint  P_\sigma (u)  P_\sigma (v) e^{-iu\hat{p}} e^{iv\hat{q}} \rho e^{-iv\hat{q}} e^{iu\hat{p}}du dv,
\end{equation}
where $P_\sigma (x)=\frac{1}{\sqrt{2\pi \sigma^2}}e^{-\frac{x^2}{2\sigma^2}}$ is Gaussian distribution function with variance $\sigma^2$.

Note that this error channel is the result after using the Pauli twirling approximation, where the coherent displacement errors are replaced by the incoherent mixture of displacement errors. One is unable to distinguish  between the pure state $|\tilde{\psi}\rangle$ and the mixed state $\mathcal{N}(|\bar{\psi}\rangle\langle\bar{\psi}|)$ by only measuring $\hat{q}$ or $\hat{p}$.
 The incoherent mixture of displacement errors will simplify our analysis but is noisier than the coherent displacement errors. The Pauli twirling approximation also  increases the average photon number
of a GKP state.

To correct small Gaussian shift errors in $\hat{p}$ and $\hat{q}$ quadratures, the Steane type error correction scheme is required. Fig.$\,${\ref{fig1}} shows the quantum circuit of the Steane type error correction, which is composed of CNOT gates, homodyne measurements and ideal GKP ancilla qubits. Many works under the circuit-based error model consider inverse-CNOT gates \cite{vuillot2019quantum,noh2020fault,noh2021low}. Nevertheless, the circuit in Fig.$\,${\ref{fig1}} doesn't involve inverse-CNOT gates, since the CNOT gate and inverse- CNOT gate are equivalent in our following discussion. In Section \ref{pre}, we will discuss the complete process of the Steane type error correction.

\subsection{The GKP code concatenated with color code}
\begin{figure}
\centering
\subfigure[]{\label{fig2a}
\includegraphics[width=3.7cm]{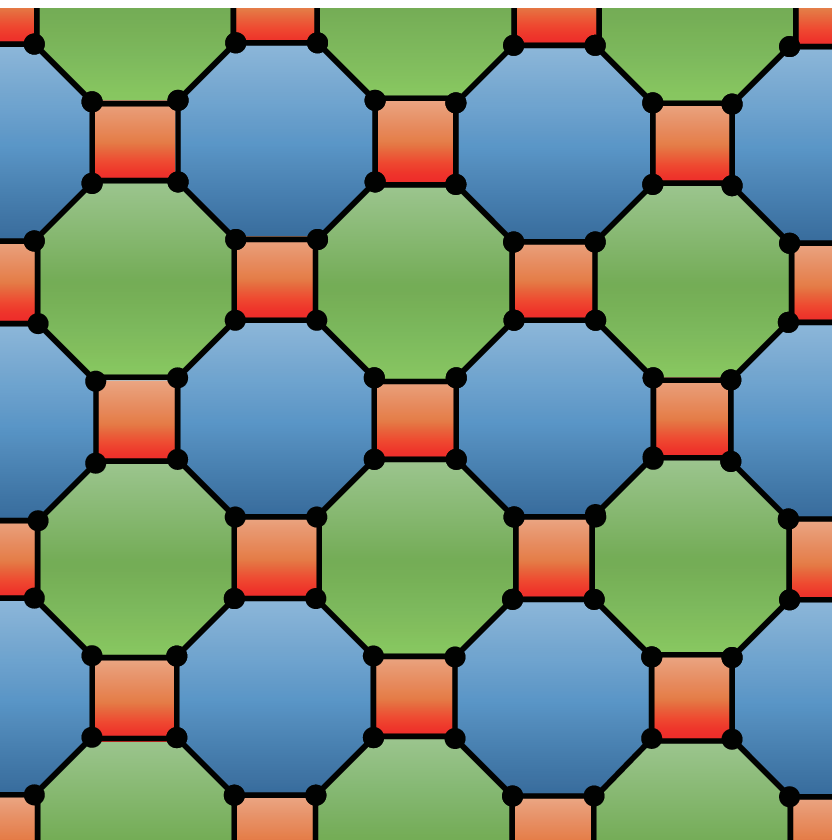}}
\hspace{0.1in}
\subfigure[]{\label{fig2b}
\includegraphics[width=3.7cm]{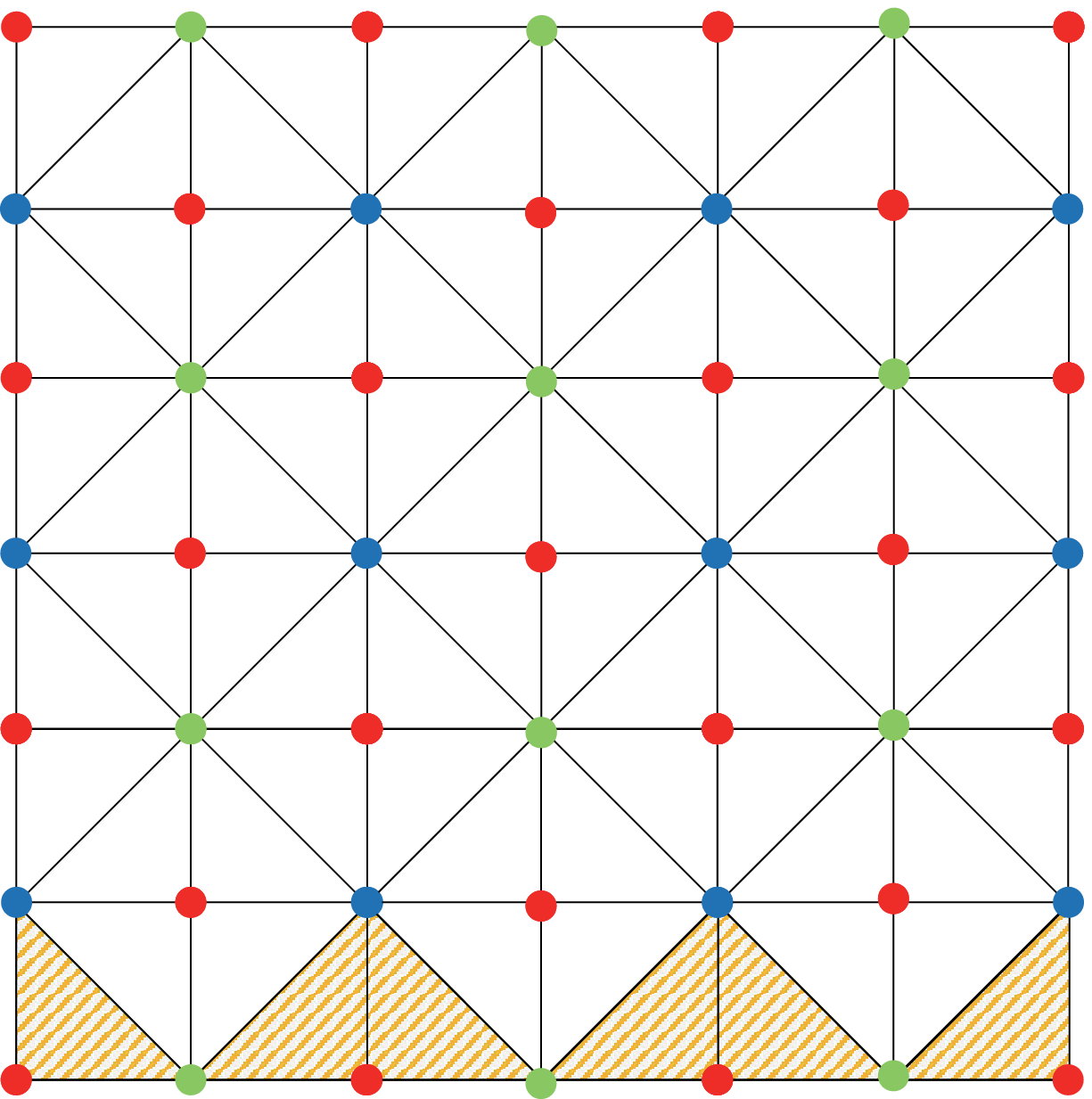}}  
\caption{\justify (Color online) The 8,8,4 color code and its dual lattice. (a) The qubit layout  of the  8,4,4 color code. Black dots indicate the data qubits and red, blue, green faces indicate syndrome qubits. The boundary faces are not complete since the whole graph embeds on the torus.
(b) The dual lattice of the color code on a torus with the parameters $\llbracket{72,4,6}\rrbracket$. In the dual lattice, the triangular faces are data qubits and the vertices are syndrome qubits. We also label a logical error of the color code which is the $X$ (or $Z$) tensor product of the shaded (yellow) qubits.}\label{f2}
\end{figure}

Color code is a  CSS stabilizer code constructed in a three-colorable trivalent graph \cite{bombin2013introduction}. Our work considers color code on the square-octagon lattice, i.e., the 8,8,4 color code \cite{fowler2011two}. 
As shown in Fig.$\,$\ref{fig2a}, the data qubits lie on the vertices in the graph, and each face $f$ corresponds to two stabilizers $X_f$ and $Z_f$, which are the tensor products of $X$ and $Z$ operators  of each qubit incident on face $f$, respectively.

The following discussion focuses on color code in the square-octagon lattice embedded on a torus with the parameters $\llbracket 2d^2,4,d \rrbracket$. Here $d$ is the code distance and logical qubit number 4  comes from the redundancy of the stabilizers:
\begin{equation}
\begin{aligned}
&\prod_{f \in red}X_f=\prod_{f \in blue}X_f=\prod_{f \in green}X_f ,\\
&\prod_{f \in red}Z_f=\prod_{f \in blue}Z_f=\prod_{f \in green}Z_f.
\end{aligned}
\end{equation}

For decoding purposes, it is beneficial to transform the color code to the dual lattice \cite{sarvepalli2012efficient} in Fig.$\,$\ref{fig2b}. The three-color vertices represent
the stabilizers (or syndrome qubits) and the triangular faces represent data qubits. Each red vertex connects with 4 faces and each blue or green vertex connects with 8 faces, which corresponds to the Pauli weight of the stabilizer. Suppose the stabilizer checks are accurate, the Pauli error in a single data qubit will lead three connecting syndrome qubits to flip. Fig.$\,$\ref{fig2b} also shows a logical error of the 8,4,4 color code which cannot be detected by stabilizer checks.

\begin{figure}[t]
\centering
\includegraphics[width=8cm]{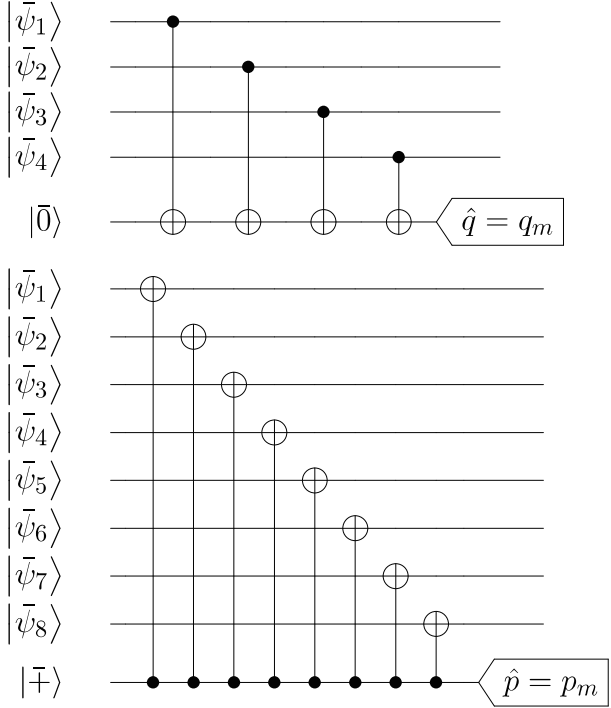} 
\caption{\justify The stabilizer check of the color-GKP code. The top is the stabilizer check of $Z_1Z_2Z_3Z_4$, and the bottom is the stabilizer check of $X_1X_2...X_7X_8$. The result of the stabilizer check depends on the measurement of the ancilla qubit. If the measurement result $q_{out}$ satisfies $|q_{out}\,{\rm mod} 2\sqrt{\pi}| <\frac{\sqrt{\pi}}{2}$, the stabilizer check is $+1$; otherwise it is -1. Note that when all the GKP states are ideal, the measurement result $q_{out}$ (or $p_{out}$) can only be the integer multiples of $\sqrt{\pi}$.}\label{fig3}
\end{figure}

Now let us consider the GKP code concatenated with the color code. The two-dimensional layout of color-GKP code follows the layout of the color code in Fig.$\,$\ref{fig2a}. Each GKP data qubit connects with ancilla GKP qubits for the GKP error correction in the first layer. And  in the second layer, these data qubits connect with syndrome qubits to implement stabilizer checks of the color code. The circuit for  stabilizer checks is presented in Fig.$\,$\ref{fig3}. Note that the $\pm$ sign of a stabilizer check depends on the $\hat{q}$ or $\hat{p}$ measurement result which is a continuous variable. If $q_{out}$ is closer to even multiples of $\sqrt{\pi}$, in other words, $|q_{out}\,{\rm mod} 2\sqrt{\pi}| <\frac{\sqrt{\pi}}{2}$, the stabilizer check is $+1$; otherwise it is $-1$.  In this paper, the range of  $a\,{\rm mod}\, b$ is from$ -\frac{b}{2}$ to $\frac{b}{2}$.

\section{ME-Steane type GKP error correction scheme}\label{pre}
Before introducing ME-Steane scheme \cite{fukui2019high}, we first simply review the conventional Steane error correction scheme. Our discussion is based on the result after Pauli twirling approximation, which means the noisy GKP state is regarded as a mixed state:
\begin{equation}\label{ee10}
\int du \int dv P_\sigma (u)  P_\sigma (v) e^{-iu\hat{p}} e^{iv\hat{q}} |\bar{\psi}\rangle\langle\bar{\psi}| e^{-iv\hat{q}} e^{iu\hat{p}}.
\end{equation} 
For the sake of simplicity, let us concentrate on the error correction in the $\hat{q}$ quadrature which needs the quantum circuit involving only one ancilla qubit, one CNOT gate and $\hat{q}$ measurement in Fig.$\,$\ref{fig1}.  Also, all components of the circuit are noiseless except the initial states. 

Assume that the input state is 
\begin{equation}
e^{-iu_1p_1} e^{-iu_2p_2}|\bar{\psi} \rangle|\bar{+} \rangle,
\end{equation}
where the probabilities of $u_1$ and $u_2$ obey the Gaussian distribution functions $P_{\sigma_1}(u_1)=\frac{1}{\sqrt{2\pi \sigma_1^2}}\exp(-\frac{u_1^2}{2\sigma_1^2})$ and $ P_{\sigma_2}(u_2)=\frac{1}{\sqrt{2\pi \sigma_2^2}}\exp(-\frac{u_2^2}{2\sigma_2^2})$, respectively. After the CNOT gate, the output state is 
\begin{equation}
\begin{aligned}
&e^{-iu_1p_1} e^{-i(u_1+u_2)p_2}|\bar{\psi} \rangle|\bar{+} \rangle\\
=&\frac{1}{\sqrt{N}}\sum_{n\in \mathbb{Z}} e^{-iu_1p_1} |\bar{\psi} \rangle|q=n\sqrt{\pi } +u_1+u_2\rangle.
\end{aligned}
\end{equation}
 Here $N$ is the normalization constant and the measurement result of the ancilla qubit is $q_{out}= n\sqrt{\pi} +u_1+u_2$. The conventional Steane  error correction scheme gives the correction $q_{cor}= q_{out}\,{\rm mod} \sqrt{\pi}$. Based on the premise that $\sigma_2$ is much smaller than $\sigma_1$, shift error $u_2$ can be ignored and then the shift error $u_1$ will be corrected successfully under the condition:
\begin{equation}
\begin{aligned}
q_{cor}=u_1+2k\sqrt{\pi},
\end{aligned}
\end{equation}
for some integer $k$. As a result, the conditional error probability for the given $q_{out}$ is
\begin{equation}\label{e9}
\begin{aligned}
p(\bar{X}|q_{out})=&1-\frac{\sum_{k}P_{\sigma_1}(q_{out}-2k\sqrt{\pi})}
{\sum_{k}P_{\sigma_1}(q_{out}-k\sqrt{\pi})}.
\end{aligned}
\end{equation}

However, when $\sigma_2$ is comparable to $\sigma_1$, the effect of the error shift $u_2$ cannot be ignored. Hence the conventional Steane scheme is not always satisfactory because it leaves the error shift $-u_2$. To find the best choice of the correction $q_{cor}$, the ME-Steane scheme  considers the distribution of the error shift $u_1$ conditioned on the measurement result $ q_{out}$ as follows:
\begin{equation}
\begin{aligned}
f(u_1,q_{out})
=\frac{P_{\sigma_1}(u_1) \sum_k P_{\sigma_2}(q_{out}-k\sqrt{\pi}-u_1)}
{\sum_k P_{\sigma_{12}}(q_{out}-k\sqrt{\pi}) },
\end{aligned}
\end{equation}
where $\sigma_{12}=\sqrt{\sigma_1^2+\sigma_2^2}$ is the variance of 
the variable $w=u_1+u_2$.  Note that $f(u_1,q_{out})$ can be decomposed into Gaussian functions with different weights:
\begin{equation}
\begin{aligned}
f(u_1,q_{out})=\frac{\sum_k
w_k
 \exp[-\frac{(u_1-\eta(q_{out}-k\sqrt{\pi}))^2}{2\sigma^2}]}
{2\pi\sigma_1\sigma_2\sum_k P_{\sigma_{12}}( q_{out}-k\sqrt{\pi})},
\end{aligned}
\end{equation}
where $\eta=\frac{\sigma_1^2}{\sigma_1^2+\sigma_2^2}$, $\sigma^2=\frac{\sigma_1^2\sigma_2^2}{\sigma_1^2+\sigma_2^2}$ and $w_k=\exp[-\frac{(\eta-\eta^2)(q_{out}-k\sqrt{\pi})^2}{2\sigma^2}]$. One can easily find the peaks of these Gaussian functions in $u_1=\eta (q_{out}-k\sqrt{\pi})$ and the maximum point of $f(u_1)$ is $u_1=\eta (q_{out}\,{\rm mod}\sqrt{\pi})$. Therefore the maximum-likelihood estimation of  shift error $u_1$ is  $q^{\scriptscriptstyle (ME)}_{cor}=\eta (q_{out}\,{\rm mod}\sqrt{\pi})$ \cite{fukui2019high}. 

\begin{figure}[t]
\centering
\includegraphics[width=8.2cm]{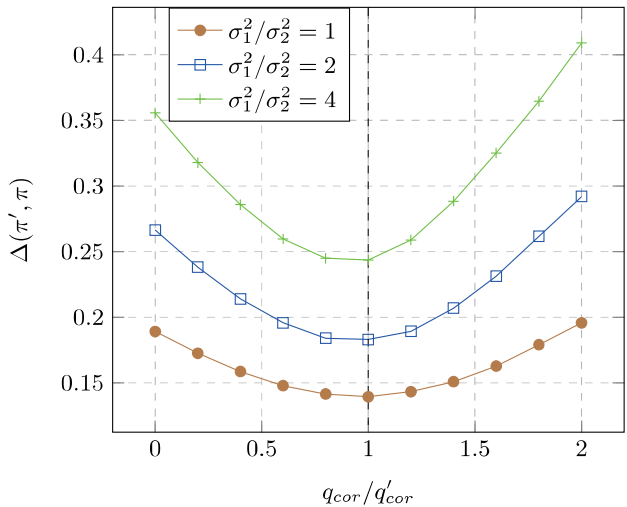} 
\caption{\justify $\Delta(\pi',\pi)$  of  the Steane type error correction with different $q_{cor}$ in Eq.$\,$(\ref{eq20}). Here $q^{\scriptscriptstyle (ME)}_{cor}=\eta ((u_1+u_2)\,{\rm mod}\sqrt{\pi})$ is the correction proposed in ME-Steane scheme. We estimate $\Delta(\pi',\pi)$ with different $q_{cor}$ by Monte Carlo simulation, where Gaussian random numbers $u_1$ and $u_2$ are produced to compute $\pi(u_1)$ and  $\pi'(u_1)$, and we repeat this process to get the average of $|(\pi'(u_1)-\pi(u_1))\,{\rm mod} 2\,\sqrt{\pi}|$.
The variance $\sigma_2^2$ of ancilla qubits is fixed as 0.056 which is the maximal squeezing of the GKP state under the current experimental technique level \cite{campagne2020quantum}. The result shows $q_{cor}=q^{\scriptscriptstyle (ME)}_{cor}$ is exactly the optimal correction to minimize $\Delta(\pi',\pi)$.}\label{fig4}
\end{figure}

In the ME-Steane scheme, the conditional $\bar{X}$ error probability is 
\begin{equation}\label{e20}
\begin{aligned}
p'(\bar{X}|q_{out})=
1-\sum_k\int_{2k\sqrt{\pi}-\frac{\sqrt{\pi}}{2}}^{2k\sqrt{\pi}+\frac{\sqrt{\pi}}{2}}f(u_1+q^{\scriptscriptstyle (ME)}_{cor},q_{out})du_1.
\end{aligned}
\end{equation}
Likewise, the conditional $\bar{X}$ error probability in   conventional scheme is 
\begin{equation}
\begin{aligned}
p(\bar{X}|q_{out})=
1-\sum_k\int_{2k\sqrt{\pi}-\frac{\sqrt{\pi}}{2}}^{2k\sqrt{\pi}+\frac{\sqrt{\pi}}{2}}f(u_1+q_{cor},q_{out})du_1,
\end{aligned}
\end{equation}
where $q_{cor}=q^{\scriptscriptstyle (ME)}_{cor}/\eta= q_{out}\,{\rm mod} \sqrt{\pi}$. 

As mentioned in the introduction, in the special case where $\sigma_2=\sigma_1$  $(\eta=\frac{1}{2})$, the correction $q^{\scriptscriptstyle (ME)}_{cor}$ is half of $q_{cor}$ \cite{noh2020encoding}. When $\sigma_2\ll\sigma_1$, we have $\eta\approx1$ and the two schemes give the same correction. In fact, the ME-Steane scheme will completely come back to the conventional scheme on the condition $\sigma_2\ll\sigma_1$. One can check that Eq.$\,$(\ref{e9}) and Eq.$\,$(\ref{e20}) give the same error probability when $\sigma_2\rightarrow 0$. Hence, in the following sections,  the conventional Steane scheme will continue to be used when assuming the GKP ancilla qubits are noiseless.

Then let us introduce a function to measure the performance of a GKP error correction scheme. A GKP error correction with perfect ancilla qubits corresponds to a mapping of the shift error $u_1$ in the data qubit:
\begin{equation}\label{e21}
\pi(u_1)=\left\{
\begin{array}{rl}
0,  & {\rm if} \ |u_1\,{\rm mod}2\sqrt{\pi}|< \frac{\sqrt{\pi}}{2};\\
\sqrt{\pi}, & {\rm otherwise}.
\end{array}
\right.
\end{equation}
This means the data qubit error $u_1$ turns into $\pi(u_1)$ after the perfect GKP error correction. Any small error $u_1$ can be corrected, and a larger $u_1$ may lead to a $\sqrt{\pi}$ displacement ($\bar{X}$ error), which will spread to the concatenated stabilizer code in the next layer. Likewise, when ancilla qubits are noisy, the mapping $\pi'(u_1)$ of a Steane type error correction scheme is
\begin{equation}\label{eq20}
\pi'(u_1)=u_1-q_{cor}.  
\end{equation}
Then we define a function $\Delta(\pi',\pi)$ as 
\begin{equation}
\begin{aligned}
\Delta(\pi',\pi)= <|(\pi'(u_1)-\pi(u_1))\,{\rm mod} 2\,\sqrt{\pi}|> ,
\end{aligned}
\end{equation}
to measure the difference between a perfect GKP error correction and a noisy GKP error correction scheme, where the angle bracket means the Gaussian weighted average of all Gaussian variables $u_1$ and $u_2$.  To show the correction $q^{\scriptscriptstyle (ME)}_{cor}$ proposed by ME-Steane scheme is  optimal, we simulate $\Delta(\pi',\pi)$ with different $q_{cor}$ and find $q^{\scriptscriptstyle (ME)}_{cor}$ is exactly the best choice to minimize $\Delta(\pi',\pi)$ (see Fig.$\,$\ref{fig4}).

One may wonder whether the maximum-likelihood estimation of shift error still remains under the coherent noise. Actually, it should be emphasized that considering the pure state as Eq.$\,$(\ref{eq7})
gives the same correction $q^{\scriptscriptstyle (ME)}_{cor}$ as above \cite{seshadreesan2021coherent}. We give a simplified proof in Appendix \ref{ab}.

\section{Decoding strategies}\label{s3}
The threshold of a quantum error correction code is largely affected
by the different error models. Therefore in this section, we first clarify two error models and then  introduce our decoding strategies of the color-GKP code. The key part of our decoding
is the MWPM algorithm which can be executed in a polynomial time.

\subsection{Error model}

\begin{figure}[t]
\centering
\includegraphics[width=8.2cm]{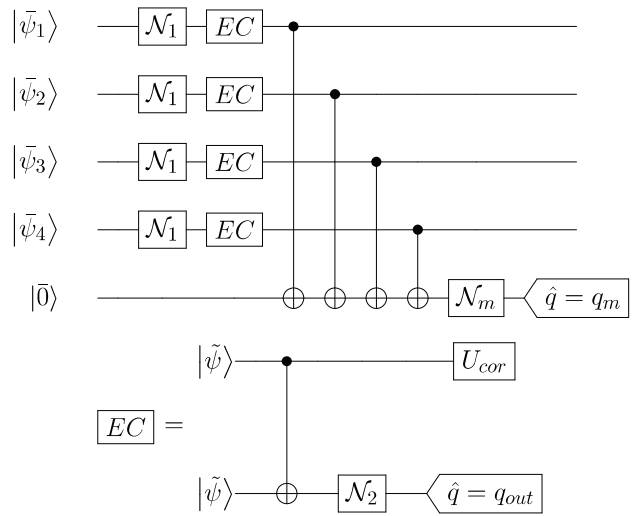} 
\caption{\justify The stabilizer check circuits with Gaussian error shift channels. The Gaussian error shift channels $ \mathcal{N}_1$, $ \mathcal{N}_2$ and $ \mathcal{N}_m$ have the same form as Eq.$\,$(\ref{ee8}). The $EC$ represents the Steane type GKP error correction with $U_{cor}=\exp(iq_{cor}\hat{p})$. The ME-Steane scheme and the conventional scheme give different $q_{cor}$ (see Section \ref{pre} for more details).}\label{fig5}
\end{figure}

In this work, all the errors are assumed to come from the Gaussian error shift channels $ \mathcal{N}_1, \mathcal{N}_2$ and $ \mathcal{N}_m$ in Fig.$\,$\ref{fig5}. Other components in the circuits are noiseless, which means the CNOT gates and the initial ancilla qubits are both perfect. Here the ancilla qubits are not only used in GKP error correction (as GKP ancilla qubits) but also in the stabilizer checks (as syndrome qubits). Moreover, after considering the effect of the Gaussian error shift channel, every measurement can be seen as the perfect homodyne measurement in $\hat{q}$ or $\hat{p}$ quadrature.

Since color code is a kind of CSS code, one can correct $X$ or $Z$ errors independently \cite{calderbank1996good,steane1996multiple}. Here we only discuss the Gaussian shift error in $\hat{q}$ quadrature and the $Z$ stabilizer check of color code.

Two kinds of error models are considered in the color-GKP error correction protocol. The first model is noisy data GKP qubits with perfect measurements. And in the second model, both data GKP qubits and measurements are noisy.

In the first error model, the measurements in GKP error correction and stabilizer checks are perfect, that is, $ \mathcal{N}_2=I$ and $\mathcal{N}_m=I$, where $I$ is the identity operator. This error model is corresponding  to the code capacity error model \cite{landahl2011fault},  in which only data qubit errors will be considered. The decoding process is implemented   after a single round of $X$ or $Z$ stabilizer checks.

It is worthy to analyze the Fig.$\,$\ref{fig5} circuit in detail. In the beginning, every data GKP qubit suffers a  Gaussian shift error $u_1$ with variance $\sigma$ from Gaussian error shift channel $\mathcal{N}_1$. Then the conventional Steane error correction is performed for every data GKP qubit. The shift $u_1$ will propagate to the GKP ancilla qubit by the CNOT gate. Therefore the measurement result of ancilla qubit is $q_{out} = u_1+n\sqrt{\pi}$, where $n$ can be any integer. Next the $\hat{q}$ correction $q_{cor}=(q_{out}\,{\rm mod} \sqrt{\pi})\in (-\frac{\sqrt{\pi}}{2},\frac{\sqrt{\pi}}{2}]$ is applied in the GKP data qubit. If 
$|u_1\,{\rm mod} 2\sqrt{\pi}| <\frac{\sqrt{\pi}}{2}$,
the correction will be successful. Otherwise, it produces odd multiples of $\sqrt{\pi}$ displacement in $\hat{q}$ quadrature, i.e., a logical $\bar{X}$ error of GKP code. 
After that, we do stabilizer checks and extract the syndromes by the output measurement results $q_{m}$. Note that since the measurements are perfect, $q_{m}$ can only be integer multiples of $\sqrt{\pi}$. So the result of a stabilizer check depends on whether $q_{m}/\sqrt{\pi}$ is odd or even.

In the second error model, in addition to the Gaussian error shift channel $\mathcal{N}_1$ in data qubits, the ancilla qubits in measurements also suffer $ \mathcal{N}_2$ and $\mathcal{N}_m$. The Gaussian error shift channel $ \mathcal{N}_2$ and $\mathcal{N}_m$ are behind the CNOT gates, so the propagation of error from ancilla qubit to data qubit won't exist, which  corresponds to the phenomenological error model \cite{dennis2002topological,landahl2011fault}. 
After applying ME-Steane error correction, every data GKP qubit is not completely corrected but carries the shift error $u_1-\eta(u_1+u_2)$, where $u_2$ is the error of the GKP ancilla qubit from $\mathcal{N}_2$. 

Since the stabilizer checks are faulty,  the stabilizer check process should repeat $d$ times typically, where $d$ is the color code distance. In each round of the stabilizer checks, we consider the effect of Gaussian error shift channels $\mathcal{N}_1, \mathcal{N}_2$ and $\mathcal{N}_m$.
Meanwhile, the measurements in the final round are assumed to be perfect both in the GKP error correction and stabilizer check. This assumption is to ensure the final GKP states are back to the code space so that one can decode the code successfully or exactly produce a logical error $X_L$ of color code. 

\subsection{The Restriction Decoder with perfect measurements}\label{s3.2}
\begin{figure*}[htbp]
\centering
\subfigure[]{
\label{fig6a}
\includegraphics[width=4.8cm]{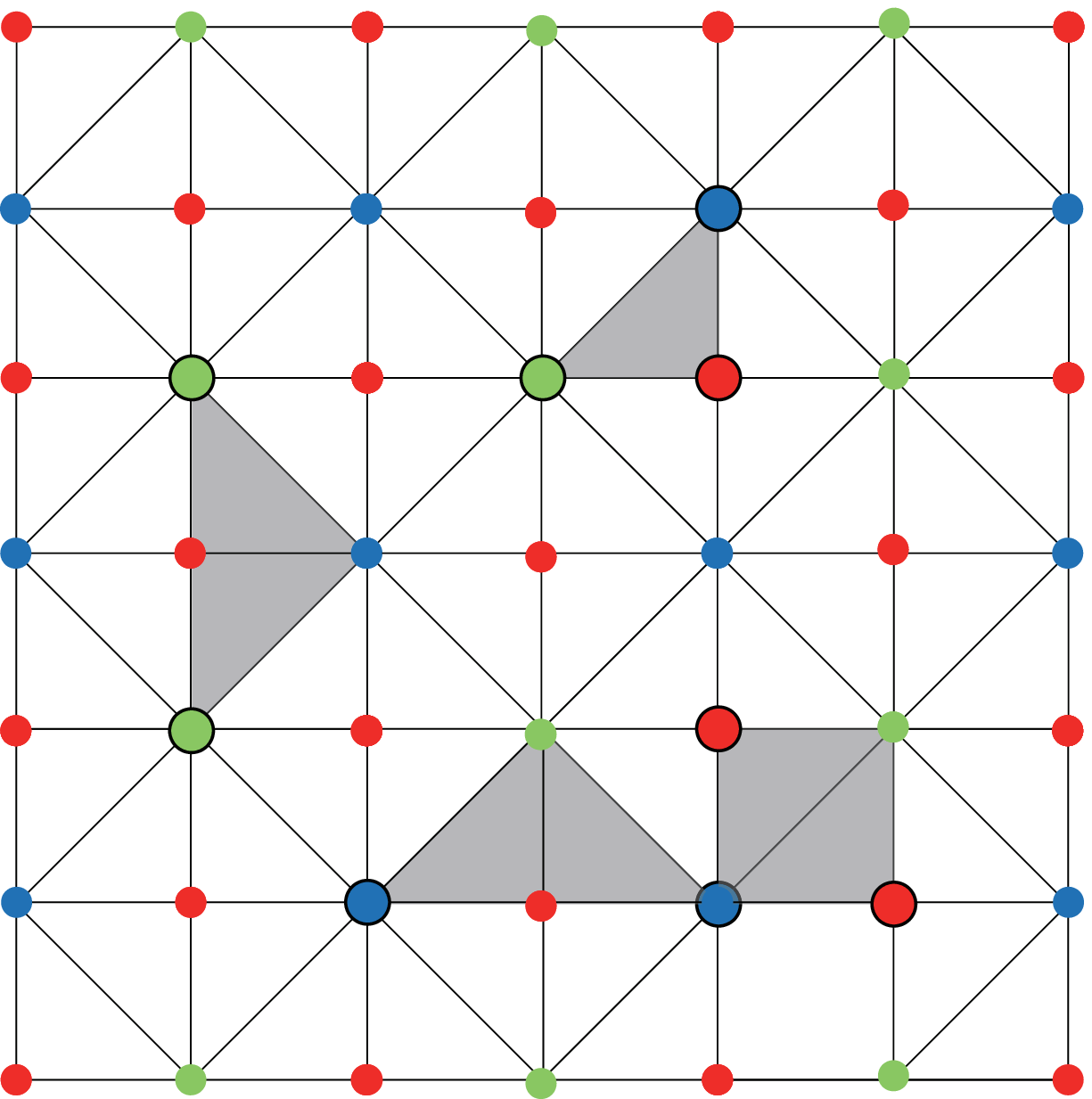}}
\hspace{0.1in}
\subfigure[]{
\label{fig6b}
\includegraphics[width=4.8cm]{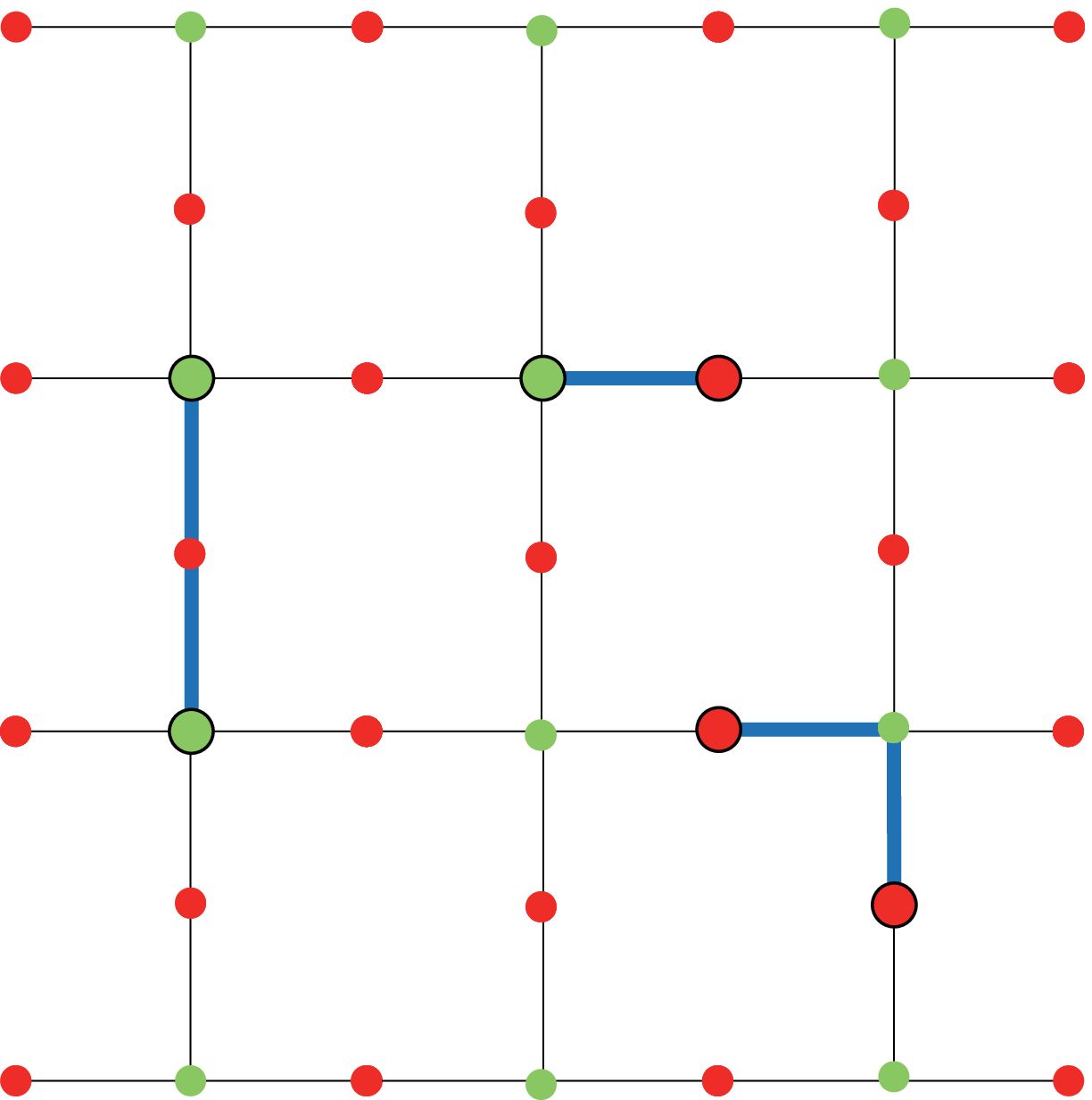}}  
\hspace{0.1in}
\subfigure[]{
\label{fig6c}
\includegraphics[width=4.8cm]{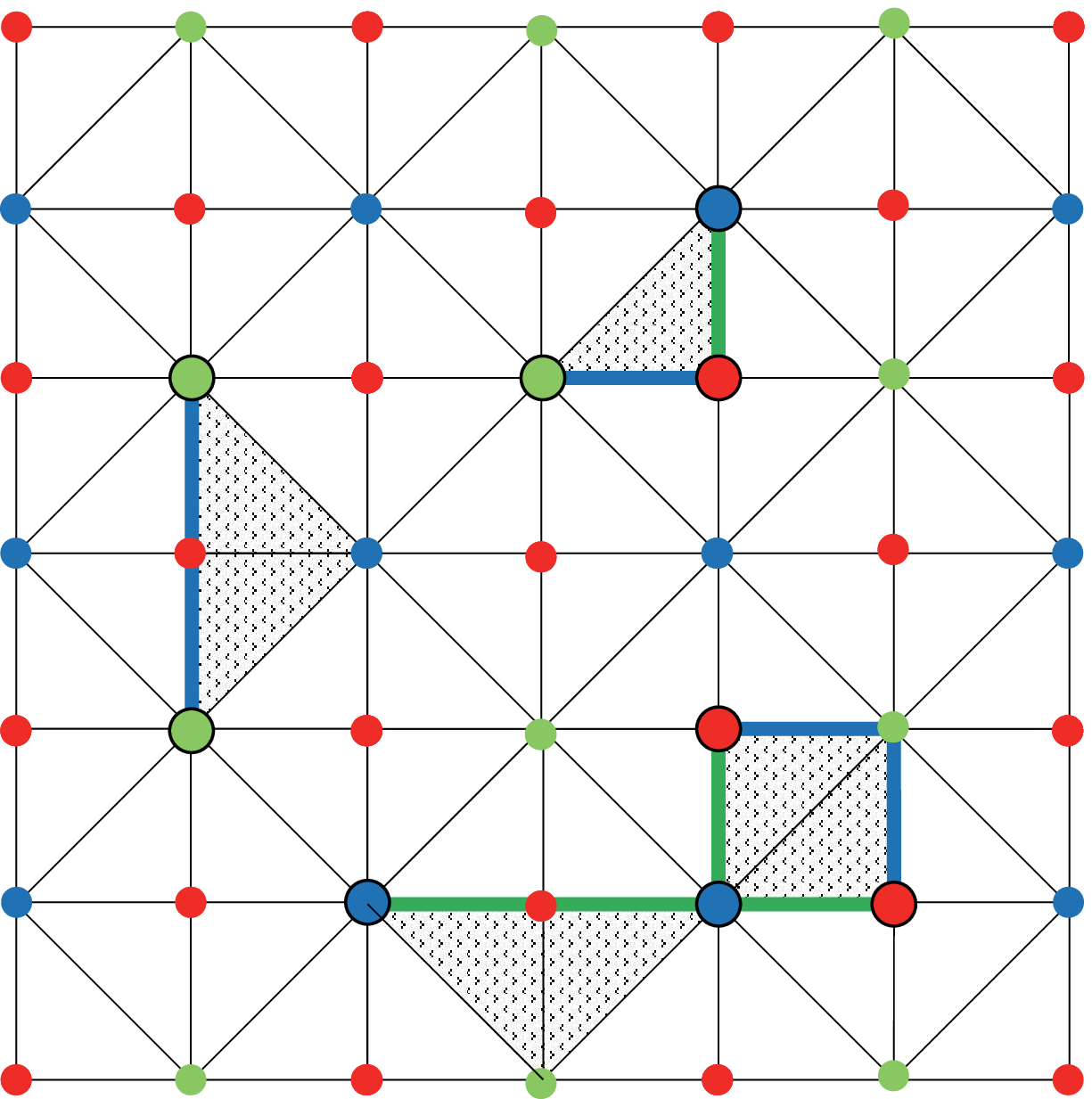}}
\caption{\justify (Color online) Restriction Decoder. (a) The error qubits (gray faces) and syndrome (amplified vertices) on the whole lattice $\mathcal{L}$. (b) The restricted lattices $\mathcal{L}_{RG}$ is obtained from $\mathcal{L}$ by removing all blue vertices, and all the edges and faces incident to the blue vertices. The bold 
(blue) lines indicate the matching results on the restricted lattices $\mathcal{L}_{RG}$. (c) Combining the matching results on $\mathcal{L}_{RG}$ and $\mathcal{L}_{RB}$ and applying the lifting procedure on each red vertex in the bold lines, one can get the correction (shaded faces). Note that the difference between corrected qubits and error qubits is just a stabilizer.
}\label{fig6}
\end{figure*}

Now we analyze the decoding of color-GKP code in the first error model.
The aim of the decoding can be described as follows:
for a given syndrome set (a set of vertices in Fig.$\,$\ref{fig6}),
the decoder finds a correction (a set of faces) that
produces the same syndromes. We hope the corrected
qubits are the same as the error qubits up to
a stabilizer, otherwise it will create a logical
error $X_L$. In general, a better decoder always finds
the qubit error which occurs with a higher
probability for fixed syndromes. The ideal decoder is the maximum-likelihood
decoder (MLD) \cite{bravyi2014efficient}. However, the 
efficient MLD for color code under the general
error model hasn't been found yet.

In our work, an efficient
decoder -- Restriction Decoder \cite{kubica2019efficient,chamberland2020triangular} will be applied , which shows a high threshold of the 8,8,4 color code by using
the MWPM algorithm in a polynomial time.  Here we decode the color-GKP
code by the Restriction Decoder combined with
the continuous variable information from the GKP
code.

The measurement results in the bosonic framework are  continuous variables rather than binary values which provides extra information
of the logical $\bar{X}$ error rate. One can compute the $\bar{X}$ error probability of GKP qubit $i$ conditioned on $q_{out,i}$ by
Eq.$\,$(\ref{e9}). In contrast, the average error probability of all GKP qubits is 
\begin{equation}
\bar{p}=1-\sum_{n\in \mathbb{Z}}\int ^{2n\sqrt{\pi}+\frac{\sqrt{\pi}}{2}}
_{2n\sqrt{\pi}-\frac{\sqrt{\pi}}{2}}P_ {\sigma}(x)dx.
\end{equation}
Apparently, using the conditional error rate rather than $\bar{p}$
is beneficial for us to find the maximum-likelihood correction.

Then let us follow the steps of the
Restriction Decoder (Fig.$\,$\ref{fig6}). First, we construct
restricted lattices $\mathcal{L}_{RB}$ and $\mathcal{L}_{RG}$, which is obtained from the whole lattice $\mathcal{L}$ by removing all green (blue) vertices of $\mathcal{L}$, and all the edges and faces incident to the removed vertices.
Using the MWPM
algorithm \cite{edmonds1965paths,higgott2021pymatching}, the syndrome vertex pairs on $\mathcal{L}_{RB}$ and $\mathcal{L}_{RG}$ can be connected. The matching results of $\mathcal{L}_{RB}$ and $\mathcal{L}_{RG}$ are two sets of edges $\rho_{RB}$ and $\rho_{RG}$. Finally,  the lifting procedure \cite{kubica2019efficient} is implemented in the edge set $\rho=\rho_{RB}\cup \rho_{RG}$
and returns the correction. Specifically, the lifting
procedure is to find all the red vertices
in edges in $\rho$ and decode the part around
each red vertex locally.

To give a proper weight using the conditional error rate, we introduce a probabilistic
matching weight for each edge (see details in Appendix \ref{ac}). Furthermore, the
shortest path connecting a syndrome vertex pair can not be  obtained
directly, for the weight is not directly related to the distance between two vertices.
So we use the Dijkstra algorithm to find
the shortest path for each possible
syndrome vertex pair before using the MWPM algorithm.

The whole decoding process can be summarized as:
  
(1) input $\sigma$, $q_{out,i}$, syndromes;

(2) compute conditional error probability $p_i$ by Eq.$\,$(\ref{e9});

(3) construct restricted lattices $\mathcal{L}_{RB}$ and $\mathcal{L}_{RG}$;

(4) use Dijkstra algorithm and MWPM algorithm to match syndrome pairs;

(5) implement the lifting procedure and output the decoding result.

\subsection{The Generalized Restriction Decoder with noisy measurements}

\begin{figure}[b]
\centering
\includegraphics[width=6.5cm]{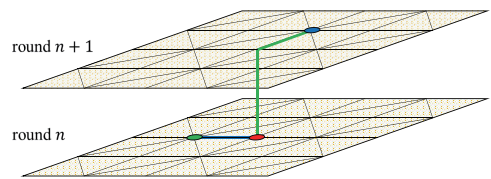} 
\caption{\justify Three-dimensional space-time graph. One data qubit error may cause syndromes in different rounds. So we match the syndrome pairs in the 3D graph and project the results to the bottom to implement lifting procedure.}
\label{fig7}
\end{figure}

Turn our attention to the second error model, where the noisy measurements exist in both GKP error corrections and stabilizer checks.  Since the results of stabilizer checks are unreliable, repeated measurements are necessary. The syndromes in different rounds can be used to construct a three-dimensional (3D) space-time graph. In the original version, the Restriction Decoder is only used to decode color code with
perfect measurements in a 2D lattice \cite{kubica2019efficient}. Another important work adapts the restriction decoder to color codes with boundaries and circuit-level noise \cite{chamberland2020triangular}.
Here we introduce the generalized Restriction Decoder to the 3d space-time graph to decode color-GKP code with noisy measurements. 

\begin{figure*}[t]
\centering
\subfigure[]{
\label{fig8a}
\includegraphics[width=8.6cm]{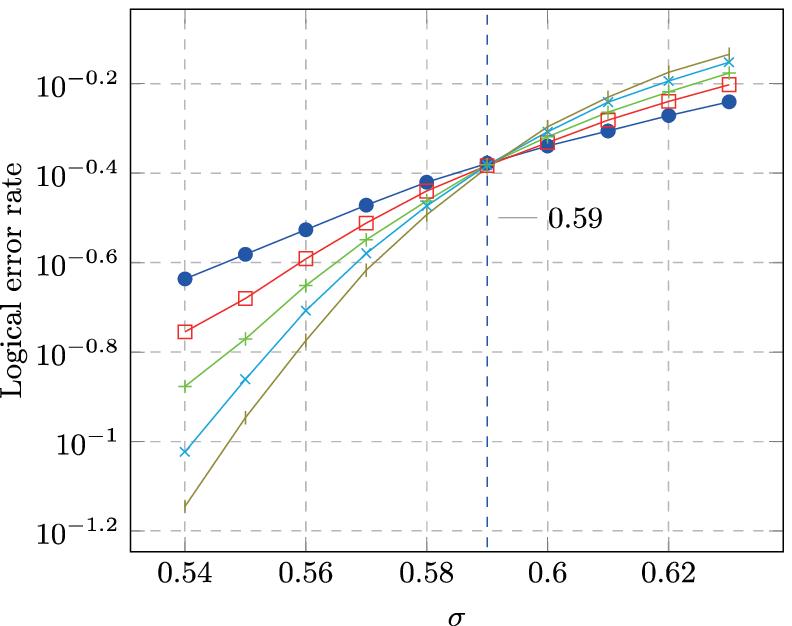}}
\hspace{0.1in}
\subfigure[]{
\label{fig8b}
\includegraphics[width=8.1cm]{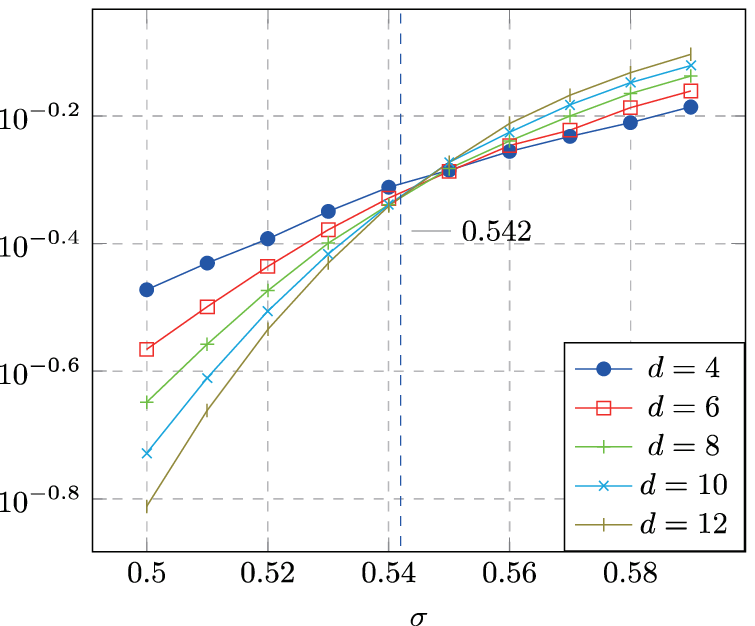}}  
\caption{\justify Threshold of color-GKP code with perfect measurements. The logical error rates are estimated by Monte Carlo simulations. We use different colors to mark the data with the different code distances $d$. (a) The threshold reaches $\sigma\approx0.59$ ($\bar{p}\approx 13.3\%$) by using the continuous-variable information.  (b) As a comparison, the threshold reaches $\sigma\approx0.542$ ($\bar{p}\approx 10.2\%$) without using the continuous-variable information. }\label{f8}
\end{figure*}

To construct the 3d cubic graph, first we extract syndromes from the measurement results $q_{m,j}$ of syndrome qubits. If  $|q_{m,j}\,{\rm mod} 2\sqrt{\pi}| <\frac{\sqrt{\pi}}{2}$, the stabilizer check is $+1$, otherwise it is $-1$. In the beginning, suppose all stabilizer checks are $+1$, then the vertices are labeled as the syndromes  whose stabilizer check is different from that in last round. For example, if  stabilizer checks in the first two rounds are 
$-1,-1$, we only label the first check since the second check doesn't change.

Next, the conditional error rates of data qubits and stabilizer checks are required. In convention Steane scheme, these conditional error rates have the expression as Eq.$\,$(\ref{e9}). The error rate $p_i$ of data qubit $i$ is obtained by just replacing $\sigma$ in Eq.$\,$(\ref{e9}) to $\sqrt{\sigma_1^2+2\sigma_2^2}$, where one $\sigma_2^2$ comes from $\mathcal{N}_2$ in the current round and the other $\sigma_2^2$ comes from last round. 
For error rate $p_{m,j}$ of stabilizer check $j$, $\sigma$  is replaced by $\sqrt{4\sigma_2^2+\sigma_m^2}$ or $\sqrt{8\sigma_2^2+\sigma_m^2}$ which depends
on Pauli weight of the
stabilizer. We also need to replace $q_{out,i}$ to measurement result $q_{m,j}$
of the syndrome qubit.

However, when taking the noise of GKP ancilla qubits into account, the data qubit error rate is given by Eq.$\,$(\ref{e20}). After the ME-Steane type GKP error correction, every data qubits  has the shift error with the variance $\eta\sigma_2^2$, where $\eta$ is not a constant in different rounds. So we replace $\sigma$ with $\sqrt{4\eta\sigma_2^2+\sigma_m^2}$ or $\sqrt{8\eta\sigma_2^2+\sigma_m^2}$ and use Eq.$\,$(\ref{e9}) to compute the error rate of the stabilizer check.

Then the Dijkstra and MWPM algorithm are performed on 3D restricted lattice in the same way of Section \ref{s3.2}. The matching weight of the horizontal
edge  follows the setting in Appendix $\,$\ref{ac}, and the weight of the vertical edge is $w_u=-\log\frac{p_{m,j}}{1-p_{m,j}}$ where $j$ is the vertex in the bottom of edge $u$. To apply the lifting procedure, we project the matching results to the bottom (see  Fig.$\,$\ref {fig7}). Note that the projections with even times in the same edge will be canceled since $\bar{X}$ error acting on one qubit twice is equivalent to the identity.

The whole decoding process can be  summarized as:
  
(1) input $\sigma_1$, $\sigma_2$, $\sigma_m$, $q_{out,i}$ and $q_{m,j}$;

(2) compute conditional error probability $p_i$ and $p_{m,j}$;

(3) construct 3d restricted lattices $\mathcal{L}_{RB}$ and $\mathcal{L}_{RG}$;

(4) use Dijkstra algorithm and MWPM algorithm to match syndrome pairs;

(5) project the matching result to the bottom;

(6) implement the lifting procedure and output the decoding result.

\section{Numerical results}\label{s4}

This section  shows our numerical simulation results of the decoding. We use the Monte Carlo simulation to estimate the logical error rates after our decoding, and determine the threshold by the common intersection point. A Gaussian random number is created in every Gaussian error shift channel $ \mathcal{N}_1$, $\mathcal{N}_2$ or $\mathcal{N}_m$ to simulate the shift errors in the circuit. It should be pointed out again that the coherent displacement errors are replaced by the incoherent mixture of displacement errors for the convenience of simulation.

The threshold of color-GKP code with perfect measurements is shown in Fig.$\,$\ref{fig8a} at $\sigma\approx 0.59$  corresponding to the average error rate $\bar{p}\approx 13.3\%$, which is close to the threshold $\sigma\approx 0.60$ of toric-GKP code in the same error model \cite{vuillot2019quantum}.  Fig.$\,$\ref{fig8b} also presents the result without using continuous variable information from the GKP code as a comparison, which shows the  threshold at $\sigma\approx 0.542$  corresponding to $\bar{p}\approx 10.2\%$. This threshold is the same as the result from the original version of Restriction Decoder where they obtained the
threshold of the normal 8,4,4 color code at  ${p}\approx 10.2\%$ \cite{kubica2019efficient}.

 \begin{figure*}[t]
\centering
\subfigure[]{
\label{fig9a}
\includegraphics[width=8.7cm]{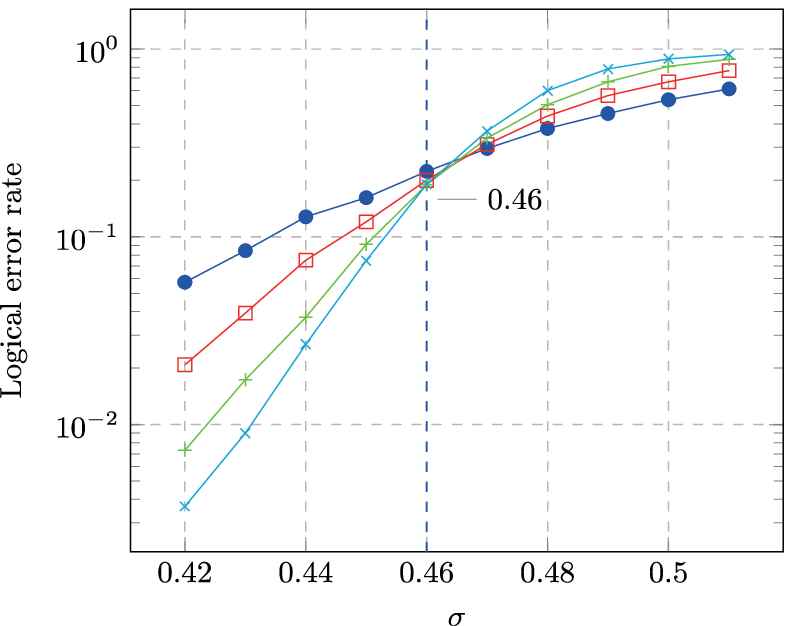}}
\hspace{0.1in}
\subfigure[]{
\label{fig9b}
\includegraphics[width=8cm]{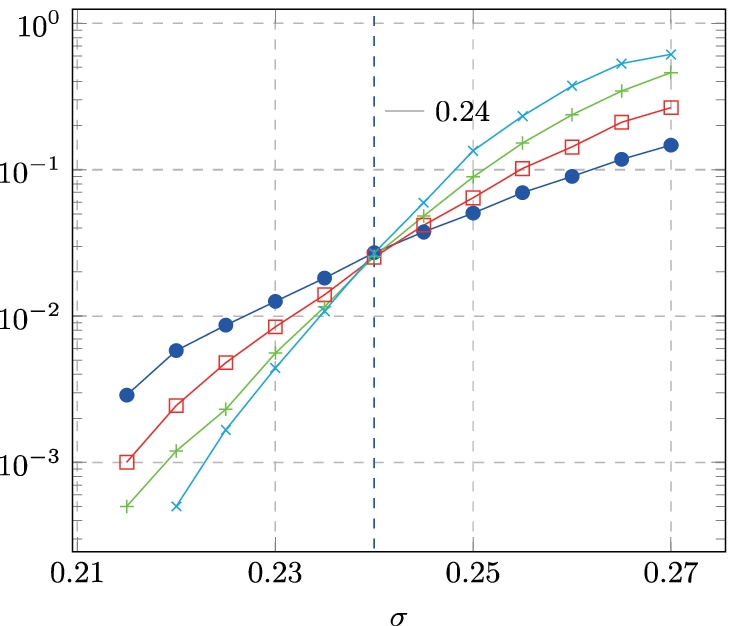}} 
\hspace{0.1in}
\subfigure[]{
\label{fig9c}
\includegraphics[width=8.7cm]{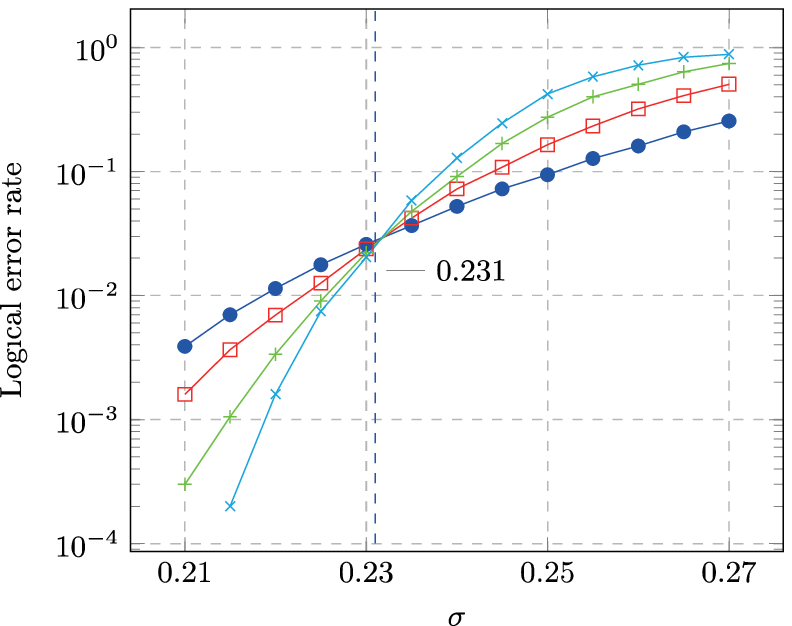}}
\hspace{0.1in}
\subfigure[]{
\label{fig9d}
\includegraphics[width=8cm]{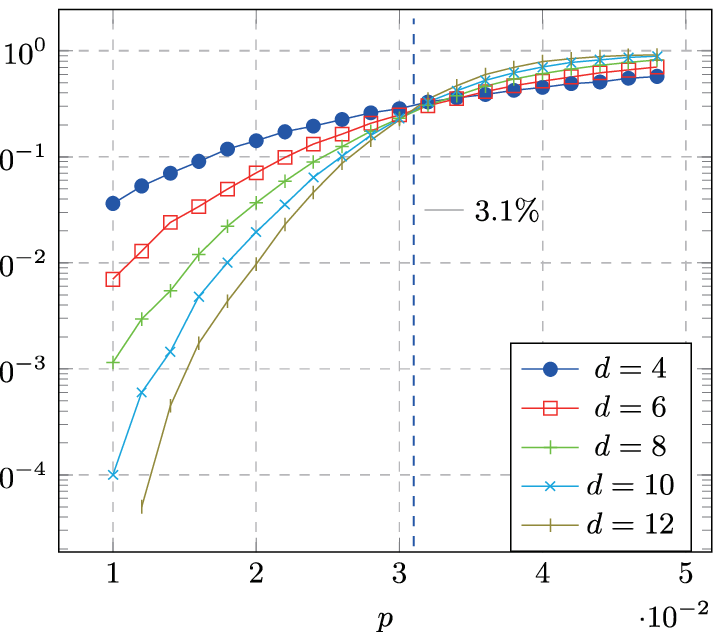}}
\caption{\justify Numerical results. (a) Threshold of color-GKP code reaches $\sigma=0.46$ with noisy stabilizer checks and perfect GKP error corrections ($\sigma_1 =\sigma_m\equiv\sigma$, $\sigma_2=0$). (b) In the case $\sigma_1 =\sigma_m=\sigma_2\equiv\sigma$, we use ME-Steane GKP error correction scheme and get the threshold at $\sigma=0.24$.  (c) Also in the case $\sigma_1 =\sigma_m=\sigma_2\equiv\sigma$, if we use the conventional Steane error correction scheme, the threshold is dropped to $\sigma=0.231$. (d) By using the generalized Restriction Decoder, the threshold of 8,4,4 color code without concatenating the GKP code reaches $p=3.1\%$ under the phenomenological error model. We use different colors to mark the data with the different code distances $d$.}
\label{fig9}
\end{figure*}

For the second error model, there are three independent noisy parameters $\sigma_1$, $\sigma_2$ and $\sigma_m$. So we just simulate two special cases: (1)
$\sigma_1 =\sigma_m\equiv\sigma$, $\sigma_2=0$; (2) $\sigma_1 =\sigma_2 = \sigma_m\equiv\sigma$. Note that the average error rates of data qubit and stabilizer measurement are equal in the first case, which corresponds to the standard phenomenological error model. The thresholds are $\sigma \approx 0.46$ in the first case and
$\sigma \approx 0.24$ in the second case (see Fig.$\,$\ref {fig9}). 

Fig.$\,$\ref {fig9b} and Fig.$\,$\ref {fig9c} also compare the ME-Steane GKP error correction scheme and  the conventional scheme in the case $\sigma_1 =\sigma_2 = \sigma_m$, which shows the threshold of color-GKP code at $\sigma\approx0.24$ and $0.231$ respectively. The distinction will be  more clear if one transforms $\sigma$ to the average $\bar{X}$ error rate $\bar{p}$. Then these two schemes give the thresholds
at $\bar{p}\approx 0.222\%$ and $0.125\%$. Note that in Fig.$\,$\ref {fig9b} and Fig.$\,$\ref {fig9c}, the logical error rates are almost equal in these two thresholds.  The consequence proves the advantage comes from the ME-Steane scheme where  it reduces about half of $\bar{X}$ errors compared with the conventional scheme. 

With the noisy measurements, the thresholds in two cases approach or even exceed
those of the toric-GKP code in the same cases \cite{vuillot2019quantum}. We attribute this good performance to the generalized Restriction Decoder in 8,8,4 color code. To demonstrate this, we test the threshold of normal 8,8,4 color code under the phenomenological error model. By using the Restriction Decoder in 3D space-time graph, the threshold reaches $p=3.1\%$ (see Fig.$\,$\ref {fig9d}).
Compared with Ref.$\,$\cite{landahl2011fault}, our decoder is efficient and gives
a higher threshold.

\section{Conclusion and discussion}\label{s5}
In this paper, we study the concatenation of GKP code with 2d color code. By applying the Restriction Decoder with continuous-variable information, the  threshold of color-GKP code is improved under perfect measurements.  Our work also generalizes the Restriction Decoder to the 3d space-time lattice to decode color- GKP code with
noisy measurements. Lastly, the good performance of the generalized decoder is also shown in normal 8,8,4 color code under the phenomenological error model.

\begin{figure}[b]
\centering
\includegraphics[width=8.2cm]{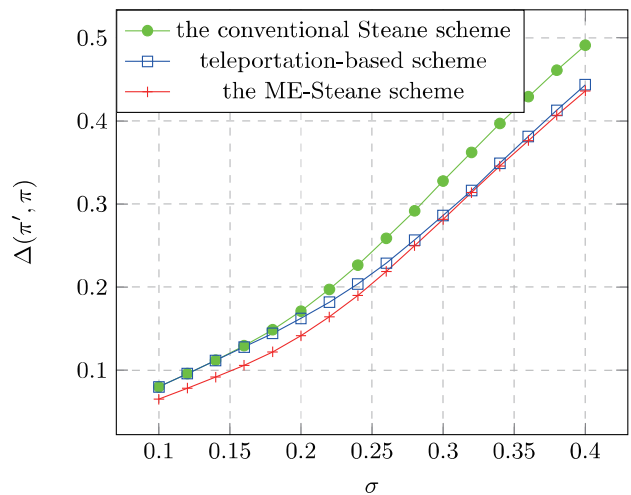} 
\caption{\justify $\Delta(\pi',\pi)$  of  three GKP error correction schemes with different $\sigma$. We assume the variance of the three initial GKP states obey $\sigma_1^2=\sigma_2^2=\sigma_3^2=\sigma^2$. The ME-Steane scheme is closer to the perfect GKP error correction than the other two schemes.}\label{fig10}
\end{figure}

We notice that Ref.$\,$\cite{walshe2020continuous} proposes the teleportation-based GKP error correction scheme using the balance beam-splitter. Ref.$\,$\cite{noh2021low} analyzes its advantages compared with the Steane type error correction scheme, where they assume the ﬁnite squeezing of the GKP states is the only noise source.
For this scheme (see more details of this scheme in Ref.$\,$\cite{noh2021low}), one can also define $\pi_t(u_1)$ and $\pi'_t(u_1)$  to characterize a perfect and a noisy GKP error correction. If GKP ancilla qubits are perfect, $\pi_t(u_1)$ has the same form as Eq.$\,$(\ref{e21}). Suppose the three initial GKP states in the teleportation have the shift error $u_1$, $u_2$, $u_3$, we have 
\begin{equation}
\pi'_t(u_1)=\frac{u_2+u_3}{\sqrt{2}}+\pi(u_1-\frac{u_2-u_3}{\sqrt{2}}),
\end{equation}
where $\pi(x)$ has been defined as Eq.$\,$(\ref{e21}).
Fig.$\,$\ref {fig10} compares $\Delta(\pi',\pi)$ (or $\Delta(\pi_t',\pi_t)$) of the Steane type schemes and teleportation-based scheme. As mentioned above, $\Delta(\pi',\pi)$ (or $\Delta(\pi_t',\pi_t)$) is a quantity to measure the difference between a perfect GKP error correction and a noisy error correction scheme. In all three schemes, we assume the shift error of the three initial GKP state $u_1$, $u_2$, $u_3$ have the same variances $\sigma^2\equiv\sigma_1^2=\sigma_2^2=\sigma_3^2$, and  other components in the circuits are noiseless.  Here $u_1$, $u_2$, $u_3$ are shift errors in $\hat{p}$ quadrature if we use the circuit in Fig.$\,${\ref{fig1}} and then 
\begin{equation}
 \pi'(u_1)=u_1-u_2-p_{cor}.
\end{equation}
We apply the noiseless error correction circuit of each scheme and simulate the error propagation. The result shows the ME-Steane scheme provides smaller $\Delta(\pi',\pi)$ than the  teleportation-based  scheme for the $\hat{p}$ error correction. Note that Steane type error correction schemes are not symmetric for $\hat{q}$ and $\hat{p}$ errors, which depends on the order of CNOT gates in Fig.$\,${\ref{fig1}}. Therefore, a better application scenarios for the ME-Steane scheme may introduce  asymmetric GKP states. We have noticed asymmetric GKP codes are 
studied in several works of the GKP concatenation code \cite{hanggli2020enhanced,noh2021low}.

Moreover, Ref.$\,$\cite{baragiola2019all} points out that only Gaussian elements are enough to obtain distillable GKP magic states. So using GKP magic states combined with transversal Clifford gates in the color-GKP code, one can theoretically achieve the universal and fault-tolerant quantum computation.

In order to connect with experiments closely, it is indispensable to consider
the circuit-level error model. The study on the surface-GKP code gives the threshold at 18.6 dB after considering the circuits in detail \cite{noh2020fault}.
But the threshold of color-GKP code under circuit-level error mode is
still unknown, which is an interesting open question for further study. Meanwhile, estimating the overhead of the color-GKP code in the realistic quantum hardware  is also another direction to extend this work. Several similar researches have been presented in the surface-GKP code \cite{noh2021low} or cat-surface code \cite{chamberland2020building}.

\acknowledgments
We thank Xiaosi Xu for the helpful discussion. This work was supported by the National Key Research and Development Program of China (Grant No. 2016YFA0301700) and the Anhui Initiative in Quantum Information Technologies (Grant No. AHY080000).

\appendix
\section{Proof of Eq.$\,$(\ref{eq7})}\label{aa}
To prove the correctness
of Eq.$\,$(\ref{eq7}) for arbitrary approximate GKP states, we only need to
show that Eq.$\,$(\ref{eq7}) holds for $|\tilde{0}\rangle$ and $|\tilde{1}\rangle$. For $|\tilde{0}\rangle$, since
\begin{equation}
\begin{aligned}
&\iint  e^{-\frac{u^2+v^2}{2\Delta^2}} e^{-iu\hat{p}}e^{iv\hat{q}} |\bar{0}\rangle du dv\\
=&\sum_{s\in \mathbb{Z}}\iint  e^{-\frac{u^2+v^2}{2\Delta^2}} e^{-iu\hat{p}}e^{iv\hat{q}} |{q=2s\sqrt{\pi}}\rangle du dv\\
=&\sum_{s\in \mathbb{Z}}\iint  e^{-\frac{u^2+v^2}{2\Delta^2}} e^{-iu\hat{p}}e^{iv2s\sqrt{\pi}} |q=2s\sqrt{\pi}\rangle du dv\\
=&\sum_{s\in \mathbb{Z}}\int  e^{-\frac{v^2}{2\Delta^2}}e^{iv2s\sqrt{\pi}}dv
\int e^{-\frac{u^2}{2\Delta^2}}  |q=2s\sqrt{\pi}+u\rangle du \\
=&\sum_{s\in \mathbb{Z}}\int  
e^{-\frac{1}{2\Delta^2}(v-2i\Delta^2s\sqrt{\pi})^2}e^{-\frac{\Delta^2}{2}(2s)^2\pi}dv\\
&\quad\times \int e^{-\frac{(q-2s\sqrt{\pi})^2}{2\Delta^2}}|q\rangle dq \\
=&\sqrt{2\pi \Delta^2}\sum_{s\in \mathbb{Z}}\int e^{-\frac{\Delta^2}{2}(2s)^2\pi}
 e^{-\frac{(q-2s\sqrt{\pi})^2}{2\Delta^2}}|q\rangle dq \propto|\tilde{0}\rangle,
\end{aligned}
\end{equation}
we have
\begin{equation}
\begin{aligned}
|\tilde{0}\rangle\propto\iint  e^{-\frac{u^2+v^2}{2\Delta^2}} e^{-iu\hat{p}}e^{iv\hat{q}} |\bar{0}\rangle du dv.
\end{aligned}
\end{equation}
Likewise, we can prove 
\begin{equation}
\begin{aligned}
|\tilde{1}\rangle\propto\iint  e^{-\frac{u^2+v^2}{2\Delta^2}} e^{-iu\hat{p}}e^{iv\hat{q}} |\bar{1}\rangle du dv.
\end{aligned}
\end{equation}
Thus as given in Eq.(\ref{eq7}), any linear superposition of $|\tilde{0}\rangle$ and $|\tilde{1}\rangle$ follows:
\begin{equation}
\begin{aligned}
\alpha|\tilde{0}\rangle+\beta|\tilde{1}\rangle\propto\iint  e^{-\frac{u^2+v^2}{2\Delta^2}} e^{-iu\hat{p}}e^{iv\hat{q}}(\alpha|\tilde{0}\rangle+\beta|\tilde{1}) du dv.
\end{aligned}
\end{equation}

\section{The pure GKP noisy state in ME-steane error correction scheme}\label{ab}
The input state is
\begin{equation}
\begin{aligned}
\int  \sqrt{P_{\sigma_1}(u_1)}&e^{iu_1p_1}|\bar{\psi}\rangle du_1
\int  \sqrt{P_{\sigma_2}(u_2)}e^{iu_2p_2}|\bar{+}\rangle du_2\\
=\frac{1}{N}\int du_1\int& du_2  \sqrt{P_{\sigma_1}(u_1)P_{\sigma_2}(u_2)}\\
&\times e^{iu_1p_1}|\bar{\psi}\rangle\sum_k|q_2=k\sqrt{\pi}+u_2\rangle.
\end{aligned}
\end{equation}
After the CNOT gate, the state is
\begin{equation}
\begin{aligned}
\frac{1}{N}\sum_k\int du_1&\int du_2 \sqrt{P_{\sigma_1}(u_1)P_{\sigma_2}(u_2)}\\
\times &e^{iu_1p_1}|\bar{\psi}\rangle|q_2=k\sqrt{\pi}+u_1+u_2\rangle.
\end{aligned}
\end{equation}
Suppose the measurement result of the ancilla qubit is $q_{out}$, the
final state is
\begin{equation}
\begin{aligned}
&N'\sum_k\int du_1\int du_2 \sqrt{P_{\sigma_1}(u_1)P_{\sigma_2}(u_2)}e^{iu_1p_1}|\bar{\psi}\rangle\\
&\qquad \times\langle q_{out}|q_2=k\sqrt{\pi}+u_1+u_2\rangle\\
=&N'\sum_k\int du_1\int du_2 \sqrt{P_{\sigma_1}(u_1)P_{\sigma_2}(u_2)}\\
&\qquad \times
\delta(q_{out}-k\sqrt{\pi}-u_1-u_2)e^{iu_1p_1}|\bar{\psi}
\rangle\\
=&N'\int du_1 \sum_k\sqrt{P_{\sigma_1}(u_1)P_{\sigma_2}(q_{out}-k\sqrt{\pi}-u_1)}e^{iu_1p_1}|\bar{\psi}
\rangle\\
\propto&\int du_1 \sum_k \exp[-\frac{u_1^2}{4\sigma_1^2}-\frac{(u_1+k\sqrt{\pi}-q_{out})^2}{4\sigma_2^2}]
e^{iu_1p_1}
|\bar{\psi}\rangle.
\end{aligned}
\end{equation}
Therefore, the probability of obtaining error $u_1$ is proportional to
\begin{equation}
\begin{aligned}
g(u_1)=|\sum_k \exp[-\frac{u_1^2}{4\sigma_1^2}-\frac{(u_1+k\sqrt{\pi}-q_{out})^2}{4\sigma_2^2}]|^2.
\end{aligned}
\end{equation}
The derivation of $g(u_1)$ is 
\begin{equation}
\begin{aligned}
\frac{{d}g}{{d}u_1}=&-\sum_k \exp[-\frac{u_1^2}{4\sigma_1^2}-\frac{(u_1+k\sqrt{\pi}-q_{out})^2}{4\sigma_2^2}]\\
&\times\sum_k 
\{(\frac{u_1}{\sigma_1^2}+\frac{u_1+k\sqrt{\pi}-q_{out}}
{\sigma_2^2})\\
&\times\exp[-\frac{u_1^2}{4\sigma_1^2}-\frac{(u_1+k\sqrt{\pi}-q_{out})^2}{4\sigma_2^2}]\}\\
=&-\sum_k \{\exp[-\frac{(u_1-\eta v_k)^2}{4\sigma^2}-\frac{(\eta-\eta^2)v_k^2}{4\sigma^2}]\}\\
\times\sum_k &
\{\frac{u_1-\eta v_k}{\sigma^2}\exp[-\frac{(u_1-\eta v_k)^2+(\eta-\eta^2)v_k^2}{4\sigma^2}]\},
\end{aligned}
\end{equation}
where $\eta$ and $\sigma$ are defined in the main body text and $v_k=q_{out}-k\sqrt{\pi}$.

Let $u_1=\eta(q_{out}-n\sqrt{\pi})$, we have
\begin{equation}\label{b6}
\begin{aligned}
\frac{{d}g}{{d}u_1}
=&-\sum_k \exp[-\frac{\eta^2(k-n)^2\pi+(\eta-\eta^2)v_k^2}{4\sigma^2}]\times\\
\sum_k &
\{\frac{(k-n)\sqrt{\pi}}{\sigma^2}\exp[-\frac{\eta^2(k-n)^2\pi+(\eta-\eta^2)v_k^2}{4\sigma^2}]\}.
\end{aligned}
\end{equation}
Since $\sigma^2\ll\pi$ is a reasonable assumption in our discussion, one
can check the exponential terms in Eq.$\,$(\ref{b6}) approximate $0$ when $k\neq n$. Therefore $u_1=\eta(q_{out}-n\sqrt{\pi})$ is the extreme value point   and $u_1=\eta(q_{out}\,{\rm mod}\sqrt{\pi})$ is the maximum point
of $g(u_1)$ if $\sigma^2\ll\pi$. This proves that the correction $q^{\scriptscriptstyle (ME)}_{cor}=\eta(q_{out}\,{\rm mod}\sqrt{\pi})$ is also a maximum-likelihood estimation for a pure noisy GKP state.

\section{The matching weight}\label{ac}
\begin{figure}[b]
\centering
\includegraphics[width=5cm]{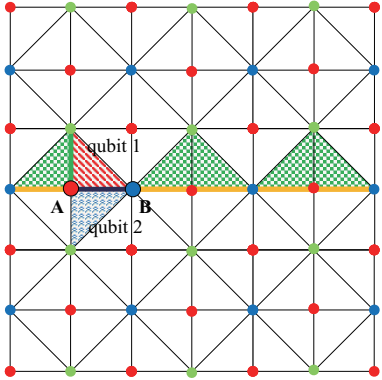}     
\caption{\justify (Color online) An example of matching weight selection. Suppose $\mathcal{L}_{RG}$ has been matched (green path) and now we are matching the vertices $A$ and $B$ on the $\mathcal{L}_{RB}$, we need to choose the black path or the yellow path as the result. We want the weight of the black (yellow) path is only related to the red (green) qubit(s). However, the weight of the black path in Eq.$\,$(\ref{ec11}) is also influenced by qubit $2$. If the error rate of qubit $1$ is small but that of qubit $2$ is huge, the matching algorithm still selects the black path. But actually, we know the error cannot come from qubit $2$ base on the result on  $\mathcal{L}_{RG}$.
}\label{fig11}
\end{figure}
 In the
original version of Restriction Decoder, the
edges with unit length (the side length of the
smallest square in Fig.$\,$\ref{fig11}) have the same
weight since every data qubit has the same error
rate. Hence a long path always has a higher
weight than a short path. But things are different
in color-GKP code where every GKP qubit has a different
conditional error probability.

Consider a path $E$ on restricted lattice $\mathcal{L}_{RB}$ or $\mathcal{L}_{RG}$. The probability of a pair of vertices connected by
 $E$ is
\begin{equation}\label{ec11}
\begin{aligned}
p_E&=\prod_{u\not \in E}(1-p_{1u})(1-p_{2u})\\
&\times\prod_{u \in E}(p_{1u}(1-p_{2u})+p_{2u}(1-p_{1u}))\\
&=P_0\prod_{u \in E}(\frac{p_{1u}}{1-p_{1u}}+\frac{p_{2u}}{1-p_{2u}}),
\end{aligned}
\end{equation}
where $u$ is the edge with unit length and $p_{1u}$, $p_{2u}$ are the conditional error rate of the data qubits adjacent to $u$. And
$P_0=\prod_{u}(1-p_{1u})(1-p_{2u})$ is a constant irrelevant to $E$.

It seems that the matching weight for edge
$u$ should be set as $w_u=-\log(\frac{p_{1u}}{1-p_{1u}}+\frac{p_{2u}}{1-p_{2u}})$. However,
this is not a good choice since only one qubit will truly affect $w_u$ but we count the probabilities of two qubits connecting with $u$. If  the matching result on one restricted lattice has been determined, then $w_u$ in the other restricted lattice is only related to one data qubit.  An example to explain this
is given in Fig.$\,$\ref{fig11}.

A better strategy is to adopt different weights on two restricted lattices according to the match order.  In the second matching (say on the  $\mathcal{L}_{RG}$), based on the known result on $\mathcal{L}_{RB}$,  the weight on restricted lattice $\mathcal{L}_{RG}$ is determined by only one qubit (say $1u$), then we set $w_u=-\log\frac{p_{1u}}{1-p_{1u}}$. In the first matching  on the  $\mathcal{L}_{RB}$, the weight $w_u$ is not deterministically affected by only $p_{1u}$ or $p_{2u}$. 
Suppose one edge $u$
is in path $E$, it may result from the error
of qubit $1u$ or qubit $2u$. For the given syndrome, the probabilities of the
former case and the latter case are
\begin{equation}
\begin{aligned}
&P_1=\frac
{p_1^{(syn)} (1-p_2^{(syn)})}
{ p_1^{(syn)} (1-p_2^{(syn)})+ p_2^{(syn)} (1-p_1^{(syn)})},\\
&P_2=\frac
{p_2^{(syn)} (1-p_1^{(syn)})}
{ p_1^{(syn)} (1-p_2^{(syn)})+ p_2^{(syn)} (1-p_1^{(syn)})},
\end{aligned}
\end{equation}
where $p_i^{(syn)}$ is the error probability of qubit $i$
conditioned on the given syndrome. To solve
$p_i^{(syn)}$ is difficult, so we do a simple
approximation where $p_i^{(syn)}$ is replaced by the qubit error
rate $p_i$.
Therefore a probabilistic matching weight is introduced for each
edge:
\begin{equation}\label{e14}
\begin{aligned}
w_u=&-\log(P_1\frac{p_{1u}}{1-p_{1u}}+P_2\frac{p_{2u}}{1-p_{2u}})\\
=&-\log[(\frac{p_{1u}}{1-p_{1u}})^2+(\frac{p_{1u}}{1-p_{1u}})^2]\\
&+\log(\frac{p_{1u}}{1-p_{1u}}+\frac{p_{2u}}{1-p_{2u}}).
\end{aligned}
\end{equation}

Actually, we do two matches on both restricted lattices. In the first match, we use the probabilistic matching weight as Eq.$\,$(\ref{e14}) for one restricted lattice. Once we know the match result in the first lattice, the matching weight for the other lattice is $w_u=-\log\frac{p_{u}}{1-p_{u}}$. After that, we exchange the matching order of $\mathcal{L}_{RB}$ and $\mathcal{L}_{RG}$ and repeat  the decoding process, and then we select the decoding result with higher probability.
\newpage
\bibliographystyle{unsrt}
\bibliography{gkp}

\end{document}